\documentclass[acmlarge]{acmart}
\AtBeginDocument{%
  }

\setcopyright{acmlicensed}
\copyrightyear{2018}
\acmYear{2018}
\acmDOI{XXXXXXX.XXXXXXX}

\acmJournal{POMACS}
\acmVolume{37}
\acmNumber{4}
\acmArticle{111}
\acmMonth{8}

\usepackage{graphicx}
\usepackage{amsthm}
\usepackage{booktabs}
\usepackage{algorithm}
\usepackage{algorithmic}
\usepackage[switch]{lineno}
\usepackage{xcolor}
 \usepackage{multirow} 
 \usepackage{threeparttable}
 \usepackage{array}
 \usepackage{tabularx}
 \usepackage{blindtext}
 \usepackage{booktabs}   
\usepackage{tabularx}   
\usepackage{array} 
\newcolumntype{Y}{>{\raggedright\arraybackslash}X}
\usepackage{longtable}
\usepackage{makecell}

\begin{document}

\title{LLMs for Human Mobility: Opportunities, Challenges, and Future Directions}

\author{Jie Gao}

\email{j.gao-1@tudelft.nl}
\affiliation{%
  \institution{Delft University of Technology}
  \city{Delft}
  \country{Netherlands}
}
\author{Yaoxin Wu}
\email{y.wu2@tue.nl}
\authornotemark[1]
\affiliation{%
  \institution{Eindhoven University of Technology}
  \city{Eindhoven}
  \country{Netherlands}}
\renewcommand{\shortauthors}{Jie et al.}

\begin{abstract}
Human mobility studies how people move among meaningful places over time and how these movements aggregate into population-level patterns that shape accessibility, congestion, emissions, and public health. Large language models (LLMs) are increasingly used in this domain because many human mobility problems require reasoning about place and activity semantics, travelers’ intentions and preferences, and diverse real-world constraints that are difficult to capture using coordinates and other purely numerical attributes. Despite rapid growth, the literature is still scattered, and there is no clear overview that connects human mobility tasks, challenges, and LLM designs in a consistent way. This survey therefore provides a comprehensive synthesis of LLM-based research on human mobility across five tasks, including travel itinerary planning, trajectory generation, mobility simulation, mobility prediction, and mobility semantics and understanding. For each task, we review representative work, connect core challenges to the specific roles of LLMs, and summarize typical LLM-based solution designs. We conclude with open challenges and research directions toward reliable, grounded and privacy-aware LLM-based approaches for human mobility. 
\end{abstract}




\maketitle



\section{Introduction}
Human mobility research investigates how individuals move among meaningful locations over time and how these personal trajectories aggregate into population-level patterns \cite{barbosa2018human,pappalardo2023future}. It goes beyond recording where and when trips occur by describing the context-driven organization of human daily life, including how long people stay in places, how often they return, and how regular routines form and vary across individuals. This individual traveler and behavior-centric view complements traditional traffic-flow models, which quantify network states such as flows, speeds, and congestion at the link or origin-destination (OD) level.  Human mobility instead asks why, when, and how individuals move between places, how heterogeneous routines and visitation frequencies arise, and how these micro-level behaviors aggregate into macro-level movement patterns. These insights provide a quantitative and qualitative basis for urban planning and traffic management, support efforts to reduce emissions and air pollution~\cite{alessandretti2020scales,luca2021survey}, and help reveal socioeconomic inequalities and spatial segregation~\cite{pappalardo2023future}.

Accordingly, an extensive body of literature has developed models and learning methods to describe, predict, and generate human mobility \cite{barbosa2018human, luca2021survey, solmaz2019survey}. Despite this progress, representing the semantic layers that drive human movement remains challenging. Latent factors such as intent, personal preferences, and situational constraints are difficult to capture using purely coordinate-based or flow-based representations. Motivated by this limitation, recent work has begun to explore large language models (LLMs) as a complementary tool for human mobility. By treating mobility as contextual activity sequences and integrating external information, LLM-based approaches offer a pathway to connect spatiotemporal traces with interpretable behavioral context. This integration promises to enhance personalization, reasoning capabilities, and long-horizon coherence across various human mobility tasks. 

Despite these opportunities, existing surveys have not yet provided a unified, task-centric synthesis of LLMs in human mobility. Current surveys often maintain a traffic-centric focus or provide only fragmented views of the field. For instance, Zhang et al.~\cite{zhang2024large} review how LLMs are combined with prior machine learning and deep learning methods for predicting traffic information and travel behaviors, and Zhang et al.~\cite{zhang2025web} present a tutorial on LLMs for traditional mobility analytics. Long et al.~\cite{long2025survey} focus on LLM applications in traffic forecasting. These works establish important foundations, but they provide limited guidance on LLM techniques tailored to human-mobility settings where modeling must account for individual routines, activity preferences, and persona-dependent constraints. In a different direction, Mayemba et al. \cite{mayemba2024short} survey mobility prediction during epidemics using machine learning, including transformer-based models and related pretrained language-model ideas; however, the scope is tied to epidemic contexts and does not generalize to the broader set of human mobility tasks. Nie et al.~\cite{nie2025exploring} propose a conceptual framework that encompasses a wide range of applications in intelligent transportation systems, such as traffic prediction, autonomous driving, and safety analytics. However, their survey primarily provides a broad overview of general transportation applications, lacking an in-depth technical analysis. Moreover, fundamental aspects of human mobility, such as travel purpose, individual preferences, and persona, are largely overlooked, despite their critical importance for understanding and modeling human mobility patterns.
Some other similar review papers on general traffic and transportation research can be found in \cite{yan2025large, hassan2025large, maksoud2025applications}. Taken together, a unified, task-centric overview that links human mobility tasks to their core challenges and to LLM roles and system designs is still missing.

This survey fills the gap by providing a task-centric synthesis of how LLMs are being adopted in human mobility research and what technical constraints shape their usefulness. We classify tasks in human mobility into five streams, namely, travel itinerary planning, trajectory generation, mobility simulation, mobility prediction, and mobility semantics and understanding. For each stream, we clarify what LLMs are actually used for, how they are integrated with other components, and why these roles are beneficial in that setting.
More broadly, the review asks, across human mobility tasks, what functional role the LLM plays and what evidence supports these choices. 
By providing a general task classification, analyzing the challenge of each task, and discussing how LLMs are applied differently to each task, this survey aims to provide practical guidance for follow-up work and inspire new research directions in the area of human mobility.

The remainder of this paper is organized as follows: Section~\ref{sec:preliminary} offers a brief background on LLMs. In Section~\ref{sec:LLMs for hm}, we classify different mobility tasks and review the literature on each of them. In Section~\ref{sec:dicsuccion}, we discuss the challenges of the tasks and the LLM opportunities in them. Finally, we conclude the work by discussing the potential future directions in Section~\ref{sec:challenge}.

\begin{figure}[t]
    \centering
    \includegraphics[width=1.0\linewidth,trim=0 3.2cm 0 3.4cm,
        clip]{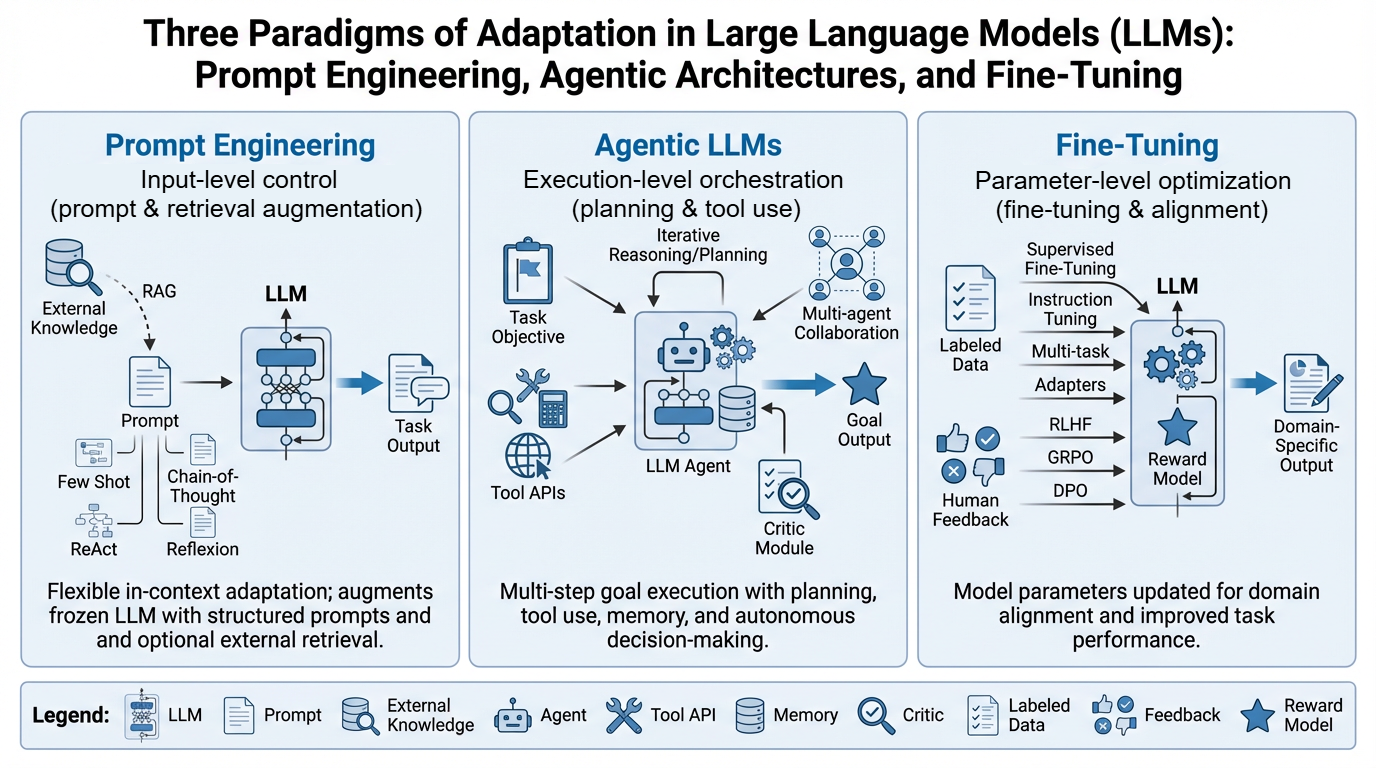}
    \caption{An illustration of three common paradigms to use LLMs.}
    \Description{A diagram showing three paradigms: prompt engineering, agentic LLM, and fine-tuning, sequentially.}
    \label{LLM3}
\end{figure}

\section{LLM Preliminaries} 
\label{sec:preliminary}
Large language models (LLMs) are neural models pretrained on large-scale corpora to predict text, most commonly through autoregressive next-token prediction. The dominant architecture is the Transformer~\citep{vaswani2017attention}, whose self-attention mechanism enables flexible conditioning on long contexts and supports strong in-context generalization.  Through pretraining at scale, LLMs acquire broad linguistic competence and a mixture of factual and procedural knowledge that can be reused for downstream tasks. In practice, LLMs are tailored to specific applications through three recurring paradigms. First, \emph{prompt-based use} leverages the model in-context, often augmented with \emph{retrieval-augmented generation} (RAG) to incorporate external, updatable, and domain-specific information at inference time.  Second, \emph{agentic architectures} orchestrate multi-step reasoning with tool use (e.g., search, solvers, simulators) and optional memory modules to support planning and iterative decision-making. Third, \emph{parameter adaptation} customizes the model via fine-tuning, including instruction tuning and parameter-efficient variants, to better match domain data, task formats, and desired behaviors. 
These paradigms appear repeatedly in the literature on human mobility, where tasks often require semantic interpretation of places and activities, integration of external knowledge (e.g., maps, timetables, POIs), and multi-step decision sequences. Fig. \ref{LLM3} illustrates the three common paradigms to use LLMs. In general, three paradigms focus on input-level, execution-level and parameter-level utilization and improvement of LLMs. For broader background on LLMs, we refer readers to \citep{naveed2025comprehensive}.  

\subsection{Prompt engineering}
Given massive and diverse text corpora, the LLMs with underlying Transformer-like architecture are trained to obtain semantic, logic and compositional patterns behind linguistic patterns, containing reasoning or task-solving abilities. Especially with the scaling of LLMs, the emergent generalization enables few-shot or even zero-shot learning capabilities in tackling different tasks. Different prompt engineering techniques have been explored. Except the zero-shot prompting, few-shot prompting~\citep{touvron2023llama}, chain-of-thought (CoT) prompting~\citep{wei2022chain}, ReAct prompting~\citep{yao2022react}, Reflexion-based prompting~\citep{shinn2023reflexion}, retrieval augmented generation (RAG)~\citep{gao2023retrieval} and others have been leveraged in different applications.

The above prompting strategies differ in how they present task structure and intermediate reasoning to the model. The few-shot prompting provides several input-output examples to guide LLM generation; CoT prompting explicitly encourages the model to generate intermediate reasoning steps, which significantly improve performance on complex reasoning tasks; ReAct prompting interacts natural language reasoning traces and tool-use actions, enabling LLMs to interact with external environments; Reflexion-based prompting introduces self-reflection mechanisms to iteratively analyze and correct LLMs' previous outputs; RAG enrich the prompt with retrieved external knowledge, which mitigates hallucination and improves factual accuracy, especially in knowledge-intensive tasks. Prompt engineering offers a lightweight and flexible paradigm to adapt frozen LLMs to downstream tasks without modifying model parameters. More details of prompt engineering can be found in~\citep{sahoo2024systematic}. 
In human mobility applications, prompt engineering is often used to involve trajectories, check-ins, or survey records into contextual natural-language descriptions. Some historical human behaviors or retrieved external resources (such as public events, sociodemographic statistics) can also be included to facilitate more accurate context learning ~\citep{liu2024human,bhandari2024urban}.

\subsection{Agentic LLMs}
Agentic LLM architectures move from single-pass generation toward multi-step goal execution.
In contrast to prompting LLMs to generate texts, agentic LLMs aim to integrate planning, tool use, and autonomous decision-making into a unified framework~\citep{schick2023toolformer,zhu2025knowagent,zeng2024perceive,hoang2025causalplan,webb2023improving}. In this regard, LLMs are designed to act as agents capable of iterative reasoning, memory utilization,  integration of external resources and adaptive behavior over multiple steps toward a goal. Recent research explores multi-agent collaboration and autonomous orchestration, where multiple LLM-based agents coordinate to perform complex tasks, often through structured communication protocols or planning hierarchies~\citep{wu2024autogen,hong2023metagpt,tao2024magis}.

There are some platforms such as AutoGPT, LangChain, and Voyager, which can be used to deploy systems of agentic LLMs, for solving complex tasks by invoking external tools (e.g., search engines, code interpreters), and evaluating progress toward long-term goals. Agentic LLMs address the limitations of prompt engineering by iteratively evaluating outcomes and refining prompts. At the same time, agentic techniques can mitigate hallucinations by incorporating critics or self-reflection modules into the decision loop. This agentic paradigm aligns naturally with human mobility applications, where decisions (such as planning) should be repeatedly made by LLM agent and checked by LLM-based critics to respect real-world constraints \citep{kambhampati2024llms,gundawar2024robust}.

Agentic LLMs represent a transition from instruction-based models to autonomous, goal-oriented systems that integrate natural language understanding, reasoning, and environmental interaction. While this advancement expands their potential in domains like robotics, software engineering, and scientific research, it also introduces complex challenges concerning controllability, alignment, and interpretability. A broader discussion of agentic LLMs is available in \citep{plaat2025agentic}.

\subsection{Fine-tuning}
Beyond in-context adaptation, LLMs can be customized through parameter adaptation. Supervised fine-tuning (SFT) is the most common approach, using labeled datasets to train an LLM in a supervised manner. Standard SFT involves updating all model parameters using task-specific losses. Instruction fine-tuning~\citep{wei2021finetuned} leverages datasets containing natural language instructions paired with desired outputs, enhancing the model’s ability to generalize across unseen tasks. Multi-task supervised fine-tuning~\citep{liu2024mftcoder} trains the LLM on multiple tasks simultaneously, using a weighted combination of task-specific losses. Parameter-efficient fine-tuning such as adapter-based fine-tuning, inserts and trains small  modules in the pre-trained LLMs while keeping the main parameters frozen. 

Reinforcement learning (RL) is used to optimize LLMs for objectives that are difficult to specify with explicit labels. Reinforcement Learning from Human Feedback (RLHF) has become a common approach, where a reward model trained on human feedback guides fine-tuning of the LLM via policy gradient methods such as Proximal Policy Optimization (PPO). Generalized Reinforcement Policy Optimization (GRPO), a variant of PPO that introduces a more flexible update rule by normalizing policy gradients across reward distributions. Direct Preference Optimization (DPO) eliminates the need for an explicit reward model by reformulating the RLHF objective as a supervised learning problem based directly on human preference data. This leads to simpler, more stable fine-tuning while maintaining alignment quality. These RL-based techniques complement supervised fine-tuning by aligning models with abstract, human-centric goals beyond simple label prediction. In human mobility research, fine-tuning can be employed to enhance the capability of LLMs to solve domain-specific tasks, such as improving planning decision-making performance and incorporating structured semantic representations \citep{shao2024chain,zhang2024agentic}.

\section{LLMs for Human Mobility}
\label{sec:LLMs for hm}

LLMs have recently been explored for human mobility because many mobility problems are naturally expressed as contextual sequences of activities and decisions. Compared with classical mobility models that operate mainly on numeric traces and engineered features, LLMs can directly represent and reason over unstructured descriptions of places, purposes, constraints, and preferences. Their pretrained knowledge about everyday activities and spatial semantics can serve as a useful prior, and, when combined with retrieval or external tools, LLM-based systems can be grounded in maps, timetables, and historical mobility records. These capabilities have motivated a growing body of work that uses LLMs either as stand-alone components or as modules within hybrid pipelines to support human-centric mobility tasks.

\begin{figure}[t]
    \centering    \includegraphics[width=\linewidth]{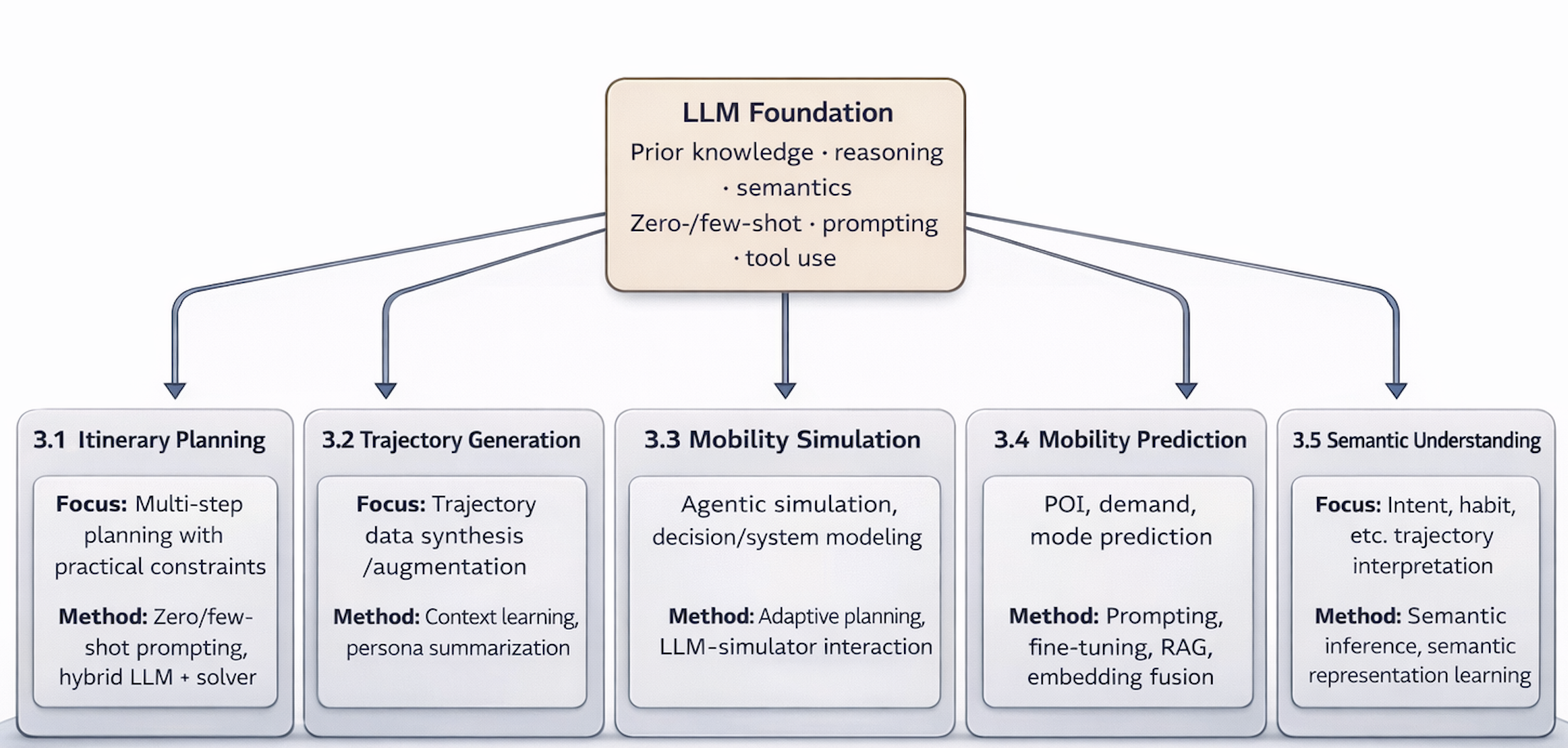}
    \caption{An Overview of LLMs for Human Mobility, where we summarize the main focus of each human mobility task along with the commonly used LLM-based techniques.}
    \Description{The picture shows the main content of our review for LLM used in different human mobility tasks.}
    \label{LLM4Mobility}
\end{figure}

In this section, we review LLM-based methods for human mobility and organize the literature by task stream. Specifically, we group existing studies into five categories, namely \textit{1) travel itinerary planning}, \textit{2) trajectory generation}, \textit{3) mobility simulation}, \textit{4) mobility prediction}, and \textit{5) mobility semantics and understanding}. 
For each category, our review is focused on how LLMs are used,
how they are integrated with other components such as retrieval, tools, or domain models, and what task-specific evidence is used to evaluate its benefits. A high-level structure of the adoption of LLMs for different human mobility tasks is shown in Fig. \ref{LLM4Mobility}. 

\subsection{LLMs for travel itinerary planning}
Travel itinerary planning aims to convert a traveler’s (often underspecified) natural language request into a feasible, time-ordered sequence of activities and trips. Recent work explores LLM-based agents for this task, where the goal is to satisfy hard constraints (e.g., opening hours, time windows, budgets, travel times, accessibility requirements, and booking availability) while improving alignment with soft preferences (e.g., interest in museums, cuisine, walking intensity). LLMs are well suited for interpreting underspecified requests and converting them into structured requirements. However, generating long-horizon itineraries that remain globally consistent is challenging when diverse constraints must be satisfied exactly and real-world facts can change. As a result, many systems integrate LLMs with external tools (such as search engines, routing services, and POI APIs) as well as verification or optimization modules to enhance reliability. 

As an early benchmark effort, \citet{xie2024travelplanner} identify key difficulties for LLM-based travel planning, including long-horizon itinerary construction, multi-dimensional hard constraints, and the need for tool use to retrieve sufficient real-world information, and they propose TravelPlanner as a benchmark reflecting realistic planning scenarios. Follow-up work extends the benchmark according to its limitations. \citet{singh2024personal} introduce a personalized variant TravelPlanner+ that incorporates user models (preferences and personal concepts) and show that personalization improves preference satisfaction compared with generic planning. \citet{banerjee2025synthtrips} propose a framework to generate queries which capture personal preferences, sustainability, and city popularity. These queries are developed from a knowledge base in \citep{banerjee2024enhancing}, and evaluated in different aspects, including groundedness with travel filters, alignment with
context, alignment with personas, adherence to sustainability filters, diversity of generated
queries, and overall clarity and quality assessment.
\citet{kambhampati2024llms} argue that reliable planning benefits from staged pipelines that refine specifications, reformulate problems, and verify candidate plans rather than relying on a single end-to-end generation step. To this end, they propose an LLM-Modulo framework in which a group of critics are used to assess the feasibility of hard and soft constraints. The critics can be model-based and LLM-based and even human-in-the-loop-based modules. The proposed approach is also applied to other planning benchmarks~\citep{valmeekam2023planbench}. \citet{gundawar2024robust} operationalize the LLM-Modulo perspective in travel planning and report that critic modules can substantially improve plan quality over common prompting baselines, although their evaluation focuses on simplified queries without rich real-world preference signals. \citet{chen2024can} empirically study commonly used LLMs and prompting/feedback strategies on TravelPlanner-style tasks and conclude that naive few-shot prompting and feedback do not reliably close the gap, while fine-tuning with feedback can improve pass rates. \citet{gadbail2025iti} propose validation rules to check the feasibility of itinerary generated by LLMs, and corresponding rules to correct the invalid itineraries. They applied the validation and correction mechanisms to the sequence of flights by retrieving the information from AeroDataBox API.

A second group of work moves beyond prompting-only planning and instead builds hybrid travel planning pipelines. Specifically, these pipelines typically use the LLM to translate free-form user requests into structured requirements and draft candidate itineraries, and then rely on formal planning or optimization tools to verify constraints and adjust the itinerary until it becomes feasible. \citet{chen2024travelagent} propose TravelAgent, a system that integrates real-time tool usage with recommendation and planning components, and uses memory to support continual personalization, reporting stronger human-evaluated rationality, comprehensiveness, and personalization across a set of travel scenarios. \citet{de2024trip} address oversubscription planning, where not all candidate POIs can be visited under time and budget limits, and propose TRIP-PAL, which uses an LLM to translate travel and user information into a planner-consumable formalization and then relies on an automated planner \citep{helmert2006new} to compute a utility-maximizing feasible plan.  \citet{hao2025large} formalize multi-constraint travel planning as constrained satisfiability and integrate sound verification tools to check feasibility and rectify plans, reporting strong gains over prompting-only agents on benchmark-style tasks. \citet{ju2024globe} integrate LLM translation with mixed-integer linear programming (MILP) to compute itineraries with precise feasibility under interdependent constraints. In this work, an LLM is used to transform user requests from natural language into symbolic format (i.e., JSON), so that MILP solver can be applied to derive the planning solution.

Benchmarks have also evolved beyond the original intercity setting and feasibility-only evaluation. \citet{shao2024chinatravel} propose ChinaTravel, an open-ended benchmark for multi-POI itinerary planning with realistic requirements and richer metrics (including preference treatment as soft constraints), and show that planning remains challenging on questionnaire-based human queries with ambiguous or undefined concepts. \citet{wang2025triptailor} introduce TripTailor with a much larger POI base and itinerary collection, and they evaluate plans along feasibility, rationality, and personalization dimensions rather than relying only on constraint pass rates.  \citet{shao2025personal} further emphasize that technical feasibility alone can produce unsatisfying itineraries. They augment benchmark settings with authentic user reviews and POI metadata to evaluate preference fit via LLM-based judging. \citet{ni2025tp} argue that prior benchmarks underemphasize spatiotemporal rationality (e.g., spatial coherence, POI attractiveness, temporal adaptability) and propose a spatiotemporal-aware benchmark where retrieval-augmented agents benefit from grounding decisions in travel experience traces.  \citet{deng2025retail} point out that real users often provide incomplete requirements and propose a decision-making support process and multi-agent refinement to elicit missing information and iteratively improve itineraries. \citet{yang2025plan} study wide-horizon planning under long context, long instruction, and long outputs and propose aspect-aware decomposition together with a simulation-based evaluation environment to assess consistency when earlier events affect later choices. \citet{tang2024itinera} introduce open-domain urban itinerary planning for “citywalk” tourism and combine LLM-based preference decomposition and retrieval with spatial optimization to improve coherence and practicality. \citet{deng2025user} further move toward dynamic, user-centric interaction by incorporating images and multimodal grounding, and propose specialized agents for destination assistance, local discovery, and disruption handling. \citet{chaudhuri2025tripcraft} make a benchmark dataset which is more practical than prior datasets. This dataset ensures geographic consistency, valid transit connectivity, and contextually accurate event and attraction information. In addition, it integrates public transit stops and schedules, and incorporates
diverse attraction and event categories. This benchmark also proposes five continuous evaluation metrics beyond prior binary validation (i.e., whether constraints are met or if one plan is better than the other).

Evaluation has started to move beyond only reporting constraint pass rates. Even when an itinerary satisfies time and budget constraints, the agent may still behave poorly, for example, by guessing missing details, failing to verify critical information, or producing untrustworthy recommendations. \citet{jiang2024towards} propose APEC-style metrics (accuracy, proactivity, efficiency, and credibility) to assess and optimize the quality of agent behavior during the planning process. In a complementary direction,  \citet{yao2025your} study robustness under fraud and misinformation in travel planning and introduce safety-oriented metrics that measure whether an agent can avoid deceptive options in the final itinerary. Table~\ref{table:llm_travel} summarizes representative work on LLM-based travel itinerary planning, including the planning setting, constraint types, datasets, and evaluation metrics.

{\footnotesize
\renewcommand{\arraystretch}{1.0}
\setlength{\tabcolsep}{4pt}

\begin{longtable}{@{}
>{\raggedright\arraybackslash}p{1cm}@{\hspace{4pt}}
>{\raggedright\arraybackslash}p{2.6cm}@{\hspace{4pt}}
>{\raggedright\arraybackslash}p{3.2cm}@{\hspace{4pt}}
>{\raggedright\arraybackslash}p{3cm}@{\hspace{4pt}}
>{\raggedright\arraybackslash}p{2.3cm}@{\hspace{4pt}}
>{\raggedright\arraybackslash}p{3.1cm}
@{}}

\caption{LLMs for travel itinerary planning. The constraint types include hard constraints (e.g. budget); commonsense constraints (e.g., minimum nights stay); environment constraints (e.g. no flights); soft constraints (i.e., historical insights such as a preference of low-cost attractions). For more details of these constraint types, readers are referred to \citep{xie2024travelplanner,chen2024travelagent}.}
\label{table:llm_travel} \\

\toprule
\textbf{Paper} & \textbf{Planning Task} & \textbf{Constraint Types} & \textbf{Role of LLM} & \textbf{Dataset} & \textbf{Evaluation} \\
\midrule
\endfirsthead

\multicolumn{6}{c}{\tablename\ \thetable\ -- continued from previous page} \\
\toprule
\textbf{Paper} & \textbf{Planning Task} & \textbf{Constraint Types} & \textbf{Role of LLM} & \textbf{Dataset} & \textbf{Evaluation} \\
\midrule
\endhead

\midrule
\multicolumn{6}{r}{Continued on next page} \\
\endfoot

\bottomrule
\endlastfoot

\cite{xie2024travelplanner} & Multi-day inter-city travel plan & Hard; Commonsense; Environment & Zero-shot planning & TravelPlanner & Constraint pass rate, delivery rate \\
\midrule

\cite{singh2024personal} & Personalized travel plan & Hard; Commonsense; Environment & Integrating preferences, zero-shot planning & TravelPlanner+ & Constraint pass/delivery rate, preference rate \\\midrule

\cite{banerjee2025synthtrips} & Travel query generation & Preference; sustainability; destination popularity & In-context learning & SynthTRIPs & Groundedness, alignment, diversity \\
\midrule

\cite{kambhampati2024llms} & Multi-day inter-city travel plan & Hard; Commonsense; Environment & Candidate plan generator, critic assistant & TravelPlanner & Constraint pass rate \\
\midrule

\cite{gundawar2024robust} & Multi-day inter-city travel plan & Hard; Commonsense; Environment & Zero-shot planning, format conversion & TravelPlanner & Constraint pass rate, delivery rate \\\midrule

\cite{chen2024can} & Multi-day inter-city travel plan & Hard; Commonsense; Environment & Zero-shot planning & TravelPlanner & Constraint pass rate, delivery rate \\\midrule

\cite{gadbail2025iti} & Inter-city flight planning & Hard; Commonsense & Zero-shot planning & AeroDataBox API & Invalid itineraries or segments, avg. issues \\\midrule

\cite{chen2024travelagent} & Personalized travel plan & Hard; Soft; Commonsense & Integrating preferences, zero-shot planning & 20 travel scenarios & Human evaluation (rationality, personalization) \\\midrule

\cite{de2024trip} & One-day single-city travel plan & Time constraints; action applicability & Problem translation (format conversion) & 100 planning tasks & Plan validity/utility, runtime \\\midrule

\cite{hao2025large} & Travel plan of flights/attractions & Hard; Commonsense & Candidate plan \,\, generation & UnsatChristmas & Constraint pass rate, delivery rate \\\midrule

\cite{ju2024globe} & Multi-day inter-city travel plan & Hard; Commonsense & Problem translation (format conversion) & TravelPlanner-like data & Pass rate, satisfaction, value, efficiency \\\midrule

\cite{shao2024chinatravel} & Multi-POI itinerary planning & Hard; Soft; Environment & Problem translation (format conversion) & ChinaTravel & Constraint pass rate, delivery rate \\\midrule

\cite{wang2025triptailor} & Personalized travel planning & Hard; Soft; Commonsense & Zero-shot planning & TripTailor & Feasibility, rationality, personalization \\\midrule

\cite{shao2025personal} & Personalized travel planning & Hard; Soft; Commonsense & Integrating preferences, zero-shot planning & RealTravel (POI metadata) & Constraint pass rate, delivery/preference rate \\\midrule

\cite{ni2025tp} & Multi-day inter-city travel plan & Spatiotemporal constraints; personalized requirements & Prompt-based RAG & TP-RAG & Spatiotemporal metric, POI semantic, query relevance \\\midrule

\cite{deng2025retail} & Multi-day inter-city travel plan & Hard; Commonsense & Agentic LLM & RETAIL & Pass rate, dialogue quality \\\midrule

\cite{yang2025plan} & One-day inter-city travel plan & User preference; POI-related and spatial constraints & Zero-shot planning & Travel-Sim & Comprehensiveness, completeness, feasibility \\\midrule

\cite{tang2024itinera} & One-day citywalk planning & Itinerary time; POI sequence & Semantic interpretation, itinerary generation & Itineraries made by travel agency & Fail rate, POI quality, etc. \\\midrule

\cite{deng2025user} & Multimodal travel planning & Constraints of mapping system & Agentic LLM & Random POIs; synthetic queries & POI detection accuracy \\\midrule

\cite{chaudhuri2025tripcraft} & Multi-day inter-city travel plan & Hard; Soft; Commonsense & Zero-shot planning & TripCraft & Temporal scores, spatial score, persona score \\\midrule

\cite{jiang2024towards} & Multi-day inter-city travel plan & Hard; Commonsense; Environment & Agentic LLM & TravelPlanner & Accuracy, proactivity, efficiency, credibility \\\midrule

\cite{yao2025your} & Travel planning with fraudulent info. & Authenticity-oriented constraints & Agentic LLM & Real-time web data & Safety-oriented metrics \\
\bottomrule

\end{longtable}}

\subsection{LLMs for trajectory generation}

Trajectory (or trip) generation aims to synthesize realistic mobility traces or activity sequences from historical observations, in contrast to itinerary planning, which constructs a single feasible plan from an explicit user request. The goal of trajectory generation is to generate synthetic trajectories whose distribution matches key properties of real mobility data, so that the outputs can be used for downstream tasks such as simulation and policy evaluation. However, mobility data is typically derived from user tracking and travel survey, which are subject to privacy concerns, participant noncompliance, and high cost. In this setting, LLMs can contribute by providing a semantic interface between raw traces and human interpretable structure.
In addition,  traditional models for trip generation only focus on generating a sequence of coordinates (i.e., locations). Instead, LLMs can be used to generate plausible natural-language rationales for a person’s movement (e.g., “going to a ramen restaurant because I’m hungry”), which can be leveraged by planners that need access to inferred motivations.
Meanwhile, traditional models often fail in the presence of major events (e.g., pandemics) due to significant distribution shifts or limited data in these exceptional situations. In contrast, LLMs can reason through a new, rare situation by its zero-shot learning capability.  

    Recent studies have started to use LLM-based agents to generate consecutive multi-day travel diaries that mimic individualized activity patterns while producing shareable synthetic data in privacy-sensitive settings. For example, \citet{li2024more} generate realistic personalized diaries from a large Shenzhen travel survey dataset. The proposed MobAgent conditions generation on individual profiles (e.g., age, income, occupation, and home/work context) and uses a two-stage pipeline: it first extracts subgroup-specific mobility patterns from profile-grouped survey trajectories with an LLM self-evaluation loop, and then generates multi-step daily activity–travel sequences through planning and recursive reasoning. The generated outputs are finally grounded to physical locations via road-network shortest-path mapping. The evaluation uses distributional similarity metrics (Jensen–Shannon divergence) over multiple statistics, including step distance, step interval, spatiotemporal OD–time patterns, and the number of daily visited locations, and reports that richer profile conditioning improves realism. Several works follow a similar “routine first, POI later” logic, where the LLM mainly ensures that the day-level narrative is coherent, and then a separate component grounds the plan in concrete locations. For instance, \citet{ju2025trajllm} propose a modular, agent-based framework, in which LLMs generate coherent daily routines (activities and intentions) and then combine them with either an LLM-based destination recommender or a mechanistic physical model to select specific POIs, which is supported by a memory module to maintain consistency over time. \citet{shao2024chain} formulate intention generation using the theory of planned behaviour~\citep{ajzen1991theory} and decompose it into daily preference prompting, anchor routine generation, and dynamic likelihood generation for the next intention. Intentions are then mapped to locations using a gravity-model-style mechanism, and the LLM is fine-tuned with prompted intentions.

Other studies treat the LLM as the main engine for extracting personal patterns and motivations from historical data before generating the diary. \citet{wang2024large} propose a two-phase framework (LLMob) in which LLMs first summarize activity patterns from persona prompts and preprocessed statistics (e.g., commuting distance and frequently visited locations) and evaluate the patterns through a self-consistency check against historical activities. The second phase captures motivations by combining recent interests/priorities with date-related motivations. Their results on a Tokyo activity dataset suggest improved realism on some temporal and routine-level measures (e.g., step interval and daily routine distribution), while other spatial metrics remain challenging. In a related line of work, \citet{liu2024human} and \citet{bhandari2024urban} generate activity chains using few-shot prompting with sociodemographic information, followed by post-processing to remove invalid outputs. However, these approaches are often limited by weak constraint modeling and by the reliance on coarse personal attributes.

A separate thread focuses on model training and constraint satisfaction. \citet{li2025geo} fine-tune Llama2 to generate trajectories that remain similar to historical traces while satisfying visit-based constraints (e.g., specific locations within specified time windows), supported by a visit-wise permutation strategy to improve infilling before and after prompted anchors. \citet{haydari2024mobilitygpt} reformulate trajectory generation as an autoregressive sequence generation task (i.e., trajectories are sequences of token), while \citet{zhang2024agentic} study fine-tuning and prompt engineering for generating diverse and consistent daily activities. \citet{du2025cams} propose a CityGPT-powered pipeline that decomposes the process into semantic routine extraction, geospatial grounding to POIs, and iterative preference alignment to improve continuity, and introduces metrics to quantify hallucination-like errors. \citet{ge2025llm} build an LLM-based agent with perception, memory, planning, reasoning, and actuation modules, emphasizing preference fit and feasibility, and evaluate distributional closeness using Kullback–Leibler (KL) divergence.

We summarize representative LLM-based trajectory generation studies, highlighting the datasets used, the generation target (e.g., diary-, POI-, or point-level traces), the scale of evaluation, and the metrics adopted to assess realism and consistency in Table~\ref{table:llm_trajectory}.

\begin{table*}[!t]
\centering
\caption{LLMs for trajectory generation.  We compare the data source, generated output (activity diary, route/link sequence, or point-level trajectory), evaluation scale, and reported metrics. Some approaches additionally generate intermediate semantic variables such as intentions or motivations before grounding outputs to locations.}
\label{table:llm_trajectory}

\footnotesize
\setlength{\tabcolsep}{2pt}
\renewcommand{\arraystretch}{1.12}

\begin{tabularx}{\textwidth}{@{}
>{\raggedright\arraybackslash}p{0.9cm}
>{\raggedright\arraybackslash}p{2.7cm}
>{\raggedright\arraybackslash}p{2.7cm}
>{\raggedright\arraybackslash}p{2.8cm}
>{\raggedright\arraybackslash}p{3.2cm}
>{\raggedright\arraybackslash}p{2.8cm}
@{}}
\toprule
\textbf{Paper} & \textbf{Dataset} & \textbf{Role of LLM} & \textbf{Generated Output} & \textbf{Scale} & \textbf{Evaluation} \\
\midrule
\cite{li2024more}  & Travel survey dataset, Shenzhen & Zero-shot generation & Individual travel diary & 25481 users; 199380 records; 2016 Nov.-2017 Jan. & JS divergence (distance, interval, OD-time, visited locations) \\\midrule
\cite{ju2025trajllm}  & Public checkin dataset, Tokyo & Zero-shot persona reasoning/generation & Individual travel diary  & Unknown & Distributional similarity, consistency metrics \\\midrule
\cite{shao2024chain}  & Tencent/mobile survey dataset & In-context learning/CoT  & Individual travel diary with intentions & 200 7-day trajectories & Statistical, semantic, aggregation evaluation \\\midrule
\cite{wang2024large}  &  Trajectory dataset, Tokyo & Zero-shot generation & Individual travel diary with motivations & 100 users; 2019 Jan.-2022 Dec. & Routine distribution metrics \\\midrule
\cite{liu2024human}  &  Travel survey dataset,
Los Angeles & Prompt-based RAG  & Individual travel diary & 264000 persons; 180000 records & JSD-based distribution similarity \\\midrule
\cite{bhandari2024urban}  & Travel survey dataset, five areas &  Zero-shot/fine-tuned generation & Individual travel diary & 10000 random
surveys for finetuning &  Pattern, trip, activity metrics \\\midrule
\cite{li2025geo}  & GPS trajectory dataset, Beijing & Prompt-based RAG & Individual travel diary with visit constraints & 182 users; 17621 trajectories;  2007 Apr.- 2012 Aug. & Distribution-similarity metrics \\\midrule
\cite{haydari2024mobilitygpt}  & GPS trajectories of taxi, Porto and Beijing & Fine-tuned generation & Generation of routes (links) & 695085 trajectories in Porto; 956070 in Beijing & JS divergence (OD, trip length, radius, gravity) \\\midrule
\cite{zhang2024agentic}  & Person trip survey data, Tokyo  & Few-shot generation & Individual travel diary & 3 groups of approximately 2500 individuals. & Spatiotemporal consistency, diversity of activity patterns \\\midrule
\cite{du2025cams}  & Tencent/ChinaMobile, Beijing & Zero-shot generation & Individual travel diary &  100000/1246 users, 297363263/4163651 trajectory points   & JSDs for  individual, collective, semantic evaluation \\\midrule
\cite{ge2025llm}  & Subway and
bus card swipe data, Beijing & In-context learning  & Sequential generation of trajectory points & 2943732 records; 65537 OD pairs; 77421 paths & KL divergence of generated and real distributions \\
\bottomrule\bottomrule
\end{tabularx}
\end{table*}

\subsection{LLMs for mobility simulation}
Another line of work uses LLMs inside mobility simulation, where many agents’ plans are executed to produce resulting movements and network-level outcomes (e.g., congestion and travel times). Unlike trajectory generation, the LLM is not used only to output a standalone diary. Instead, it can be part of an interactive loop: (i) the LLM proposes an agent’s plan or a plan update, (ii) the simulator executes the plan and returns the resulting conditions, and (iii) the LLM may revise the plan when disruptions occur or when feedback indicates that the original plan is no longer reasonable. 

For example, \citet{song2025incorporating} integrate LLMs with agent-based modeling (ABM) to simulate mobility behaviors in stages: it aggregates individuals into age-based groups, assigns start/end locations based on income and occupation, selects occasional locations using attractiveness and distance, and then determines routes using the McRAPTOR algorithm~\citep{delling2015round}. This framework represents a coupling between LLM-driven behavioral components and a mechanistic routing module. However, it lacks quantitative evaluation of the generated trajectories, limiting assessment of realism beyond plausibility.

Other frameworks adopt a hybrid compromise to handle city-scale computational cost and scalability. \citet{liu2025mobiverse} propose a pipeline, in which a non-LLM activity-chain generator first creates each agent’s daily plan (e.g., activity type, start/end time, POI), and the SUMO simulator executes these plans to produce realized trajectories and system-level outputs. The LLM is used mainly for adaptive replanning as an activity-chain modifier for affected agents under disruptions (e.g., road closures, congestion, events), with prompt engineering and parallelization to control cost. In the same spirit, \citet{yan2024opencity} focus on the system layer prompt and execution efficiency to enable large-population simulation with a standard “Generative Agent” workflow. \citet{liu2025gatsim} position the LLM as an online decision-maker inside a running simulator, using it for activity planning or plan revision and reactive behavior, which then yields agents’ movements and emergent traffic patterns. Table~\ref{table:llm_mobility_simulation} summarizes representative LLM-augmented mobility simulation frameworks and contrasts how the LLM is coupled to the simulator, the scale considered, and the evaluation protocols.

\begin{table*}[!t]
\centering
\caption{LLMs for mobility simulation.}
\label{table:llm_mobility_simulation}

\footnotesize
\setlength{\tabcolsep}{2pt}
\renewcommand{\arraystretch}{1.12}

\begin{tabularx}{\textwidth}{@{}
>{\raggedright\arraybackslash}p{0.8cm}
>{\raggedright\arraybackslash}p{2.6cm}
>{\raggedright\arraybackslash}X
>{\raggedright\arraybackslash}p{3.8cm}
>{\raggedright\arraybackslash}p{2cm}
>{\raggedright\arraybackslash}p{2.4cm}
@{}}
\toprule
\textbf{Paper} & \textbf{Simulation} & \textbf{Role of LLM} & \textbf{Coupling Mechanism} & \textbf{Scale} & \textbf{Evaluation} \\
\midrule
\cite{song2025incorporating}
& Urban mobility simulation
& Reasoning locations in human trajectories
& LLM-based activity/location generation + McRAPTOR routing
& 94770 agents
& Limited quantitative validation \\\midrule

\cite{liu2025mobiverse}
& Urban mobility simulation
& Replanning with changing environment
& Traffic simulation (SUMO) + LLM for activity replanning
& 53000 agents
& Runtime for varying agent numbers\\\midrule

\cite{yan2024opencity}
& Urban mobility simulation
& Human-like behavioral generation
& Agent behavioral generation + simulator execution
& 10000 agents
& Simulation time, faithfulness (JSD) \\\midrule

\cite{liu2025gatsim}
& Online interactive traffic simulation
& Activity planning with reflection and memory
& Agent activity planning + simulator execution
& 70 agents
& Statistical analysis and test\\
\bottomrule\bottomrule
\end{tabularx}
\end{table*}

\begin{table*}[!t]
\centering
\caption{LLMs for mobility prediction. We summarize representative work across prediction targets, reporting the functional role of the LLM, the data modality, and the evaluation metrics used in each study.}
\label{table:llm_mobility_prediction}

\footnotesize
\setlength{\tabcolsep}{2pt}
\renewcommand{\arraystretch}{1.12}

\begin{tabularx}{\textwidth}{@{}
>{\raggedright\arraybackslash}p{0.9cm}
>{\raggedright\arraybackslash}p{3.7cm}
>{\raggedright\arraybackslash}p{3.8cm}
>{\raggedright\arraybackslash}p{4.3cm}
>{\raggedright\arraybackslash}p{2.8cm}
@{}}
\toprule
\textbf{Paper} & \textbf{Task} & \textbf{Role of LLM} & \textbf{Dataset} & \textbf{Evaluation} \\
\midrule
\cite{xu2025evaluating}
&  Travel mode choice prediction
& RAG + in-context learning
& 2847 labeled trips
& Accuracy, precision, recall, F1 \\\midrule

\cite{liu2025aligning}
&  Transport mode choice prediction
& Persona-context-conditioned predictor
&  1192 travels, 9 responses to choice contexts
& JSD, F1 score \\\midrule

\cite{liu2024can}
& Travel mode choice prediction
& Few-shot learning
& 9036 responses from 1004 respondents
& Cross-entropy, F1 \\\midrule

\cite{alsaleh2025towards}
& Travel mode choice prediction
& Fine-tuning with QLoRA
& 100 respondents with responses, 3 datasets
& Accuracy, precision, recall, F1 \\\midrule

\cite{badawi2025harnessing}
&Transport mode prediction
&  Zero-shot predictor
& 4280 trips (1070 per mode)
& Accuracy, precision, recall \\\midrule

\cite{wang2023would}
& Next location prediction
& Context-conditioned zero-shot predictor
& Check-in (FSQ-NYC), GPS trajectory (Geolife)
& Accuracy, nDCG@k, weighted F1 \\\midrule

\cite{liu2024nextlocllm}
& Next location prediction
& Semantic-enhanced  coordinate regressor
& Trajectory dataset of Xian\&Chengdu\&Japan
& Hit@1, Hit@5, Hit@10 \\\midrule

\cite{feng2024agentmove}
& Next location prediction
& Zero-shot predictor
& Foursquare checkin data; ISP GPS
trajectory data
& Acc@1, Acc@5, NDCG@5 \\\midrule

\cite{li2025zero}
& Next location prediction
& Two-stage prediction (activity category → POI)
& Foursquare-NYC, Foursquare-TKY, Gowalla-CA
& Acc@1, Acc@5, Acc@10, Acc@20, MRR \\\midrule

\cite{qin2025foundational}
& Individual mobility prediction
& In-context learning, CoT, fine-tuning
& Check-in (FSQ-NYC), GPS trajectory (Geolife), metro trip (HK-ORI)
& Accuracy, weighted F1\\\midrule

\cite{gong2024mobility}
& Intent-aware mobility prediction
& In-context learning, CoT, fine-tuning
& Check-in (Gowalla, FourSquare, WeePlace,  Brightkite)
& Acc@1, Acc@5, Acc@20, MRR \\\midrule

\cite{liang2024exploring}
& Mobility prediction under \,\,public events
& In-context learning, CoT
& NYC taxi/event data
& Demand prediction error (e.g., MAE, RMSE) \\\midrule

\cite{zhong2024hmp}
& Mobility prediction under \,\,\,\,\,\,rare events
& In-context learning, CoT
& Human mobility data collected by smart devices
& RMSE, MAE, MAPE \\\midrule

\cite{yang2025causalmob}
& Mobility prediction under \,\,public events
& Event embedding extractor
& GPS trajectory during public events
& RMSE, MAE, MAPE \\\midrule

\cite{li2024limp}
& Intent-aware mobility prediction
& Agentic LLM, fine-tuning
& Mobile application location data
& Acc@1, Acc@5, Acc@10, MRR@5 \\\midrule

\cite{nie2025joint}
& Regional demand prediction
& Semantic region encoder
& Package/food delivery datasets
& MAE, RMSE \\
\bottomrule\bottomrule
\end{tabularx}
\end{table*}

\subsection{LLMs for mobility prediction}
LLM-based methods have recently been explored for a range of mobility prediction tasks, spanning largely on mode choice prediction, next location prediction and aggregate mobility prediction under disruptions. A shared motivation is that mobility datasets often contain information that is either (i) heterogeneous and weakly structured (e.g., POI semantics, trip context, textual event descriptions) or (ii) data-scarce in critical regimes (e.g., rare disasters, cross-city transfer, limited labeled choice data). LLMs are attractive in this setting because they can translate complex inputs into structured signals, exploit semantic knowledge about places and activities, and support few-shot or zero-shot prediction when conventional machine learning models struggle to generalize.
Table~\ref{table:llm_mobility_prediction} summarizes representative studies on LLM-based mobility prediction, grouped by prediction target (choice modeling, next-location ranking, event-aware forecasting, and regional demand prediction), and highlights the functional role of the LLM and the evaluation protocols used in each setting.

Specifically, one stream of work focuses on LLMs for \textbf{mode choice prediction}.
\citet{xu2025evaluating} formulate travel mode choice prediction, including Drive, Walk, Transit, Bike or Micro-mobility, as an LLM-based classification problem, where each trip record is serialized into text and the LLM predicts a mode in a few-shot manner. The key contribution is the systematic evaluation of retrieval-augmented generation (RAG) strategies: similar historical trips are retrieved from a Facebook AI similarity search (FAISS) vector database and inserted into the prompt to ground the decision in empirical precedents, enabling a controlled comparison of retrieval designs within the same LLM-based pipeline. A closely related research direction focuses on employing LLMs to predict discrete travel choices by conditioning on latent behavioral factors. \citet{liu2025aligning} predict mode choice by introducing persona-based conditioning: an expert LLM infers personas that reflect latent preference trade-offs (e.g., cost versus comfort), and a learnable loading function maps each traveler to an appropriate persona, improving robustness under limited data. \citet{liu2024can} also study mode choice prediction and report that naive zero-shot prompting is insufficient to reproduce human choice behavior; performance improves when predictions are grounded through few-shot reference cases and transferable persona-style information. Beyond prompting, \citet{alsaleh2025towards} provide a large-scale evaluation of open-access causal LLMs for mode choice and show that parameter-efficient fine-tuning (e.g., LoRA) is capable of yielding stronger classification performance. Finally, \citet{badawi2025harnessing} investigate training-free mode detection from GPS trips by converting engineered trajectory descriptors into structured natural-language descriptions and prompting an LLM to classify the mode, targeting scenarios where labeled training data is limited.

Another line of work focuses on predicting an individual’s \textbf{next location}. \citet{wang2023would} design a direct next-location prediction task and prompt an LLM using historical stays and contextual stays to reflect both long-term routine dependencies and short-term context, enabling time-aware prediction. Several papers emphasize that next-location prediction is difficult to transfer across cities when framed as discrete POI-ID classification. \citet{liu2024nextlocllm} address this by reformulating next-location prediction as coordinate regression, predicting the next destination’s coordinates and then retrieving top-$k$ nearby POIs for evaluation. An LLM is used to inject POI semantics via natural-language category embeddings and to fuse semantic and spatiotemporal signals before a lightweight regression head outputs coordinates. But the out-of-distribution coordinate prediction might induce the performance degradation. \citet{feng2024agentmove} target zero-shot next-location prediction and treat the LLM more explicitly as an agentic reasoner. It combines long/short-term memory summaries, textual location context (e.g., reverse-geocoded descriptions), and collective transition signals (tool-assisted) to output a ranked next-POI list in POI/ID space. \citet{li2025zero} further impose a hierarchical reasoning structure by decomposing the forecasting problem into two stages: an activity-level planner predicts the next activity category (e.g., “restaurant”), and then a location-level selector chooses a specific POI within that category. \citet{qin2025foundational} further target transferability by proposing MoBLLM, an open-source LLM fine-tuned for generic individual mobility prediction (e.g., next location/trip information) across multiple cities and mobility data sources, using instruction tuning to improve robustness to different prompt styles and contexts.
The key reason LLMs can help here is semantic generalization: locations are not only coordinates or IDs, but carry functional meaning (work-like, leisure-like, meal-like), and LLMs can exploit textual POI/category information to improve transfer and reduce overfitting to city-specific IDs. LLMs can also be used for multiple tasks beyond next location prediction. \citet{gong2024mobility} reformulate LLMs for three tasks on check-in sequences, including next-location prediction, next-time prediction, and trajectory user linking. Their design learns check-in representations and matches them to intention prompts, using the inferred intentions together with check-in embeddings as inputs to the LLM for downstream prediction. A central claim is that semantic intention modeling enables competitive performance in few-shot settings, reducing dependence on large labeled datasets.

Beyond individual forecasting, several papers use LLMs to predict \textbf{aggregate mobility patterns} under public events or disruptions. \citet{liang2024exploring} predict daily travel demand under public events by combining a historical-average component for regular demand with LLM-based reasoning conditioned on event descriptions and historical demand context. Chain-of-thought prompting is used to guide reasoning and provide interpretable narratives for event-driven deviations. Similarly, \citet{zhong2024hmp} adopt a decomposition strategy: an LLM first predicts normal mobility patterns and a second component corrects predictions based on disruptive events such as COVID-19. Both studies argue that LLMs can be effective in data-limited or distribution-shift regimes (events/emergencies) because they can leverage broad prior knowledge to interpret rare contexts, while also acknowledging that pure zero-shot reasoning without grounding is often insufficient for strong predictive accuracy. In a causality-driven formulation, \citet{yang2025causalmob} propose CausalMob and use an LLM to convert unstructured news into structured event descriptors and intention-like scores (e.g., danger, interest, transport impact), which then serve as treatment features for causal analysis and prediction of mobility impacts. LLMs have their ability to translate unstructured event narratives into structured predictive features and to support scenario-based reasoning when historical analogs are scarce. The limitation is that performance depends heavily on how event information is grounded and quantified (e.g., extraction quality, scoring stability, and the causal model’s assumptions).

Please note that, not all papers use LLMs as the final predictor. Some works use LLMs to generate semantic signals (e.g., intentions, preferences, region semantics) that improve conventional predictive models. \citet{li2024limp} employ an LLM to infer latent intent annotations from trajectories (e.g., working, errands, leisure) because such intent is typically unobserved. Then, the inferred intent distribution is added as an input to a separate next-POI prediction model, improving prediction accuracy by making intent explicit. At the regional level, \citet{nie2025joint} leverage an LLM as a generic geospatial encoder to extract semantic region representations from location data, construct a semantics-informed graph, and perform inductive demand prediction for new regions by leveraging observed regions during training.
In these works, the value of LLMs lies in semantic enrichment: they provide structured latent variables (intent labels/probabilities or region semantics) that are difficult to obtain directly from raw mobility logs, enabling downstream predictors to model behavior more explicitly and achieve stronger generalization.

\begin{table*}[!t]
\centering
\caption{LLMs for mobility semantics and understanding. We summarize representative work across semantic targets, reporting the functional role of the LLM, the downstream task, and the evaluation metrics used in each study.}
\label{table:llm_semantics_mobility}

\footnotesize
\setlength{\tabcolsep}{2pt}
\renewcommand{\arraystretch}{1.12}

\begin{tabularx}{\textwidth}{@{}
>{\raggedright\arraybackslash}p{0.9cm}
>{\raggedright\arraybackslash}p{4.cm}
>{\raggedright\arraybackslash}p{3.4cm}
>{\raggedright\arraybackslash}p{3.9cm}
>{\raggedright\arraybackslash}p{2.9cm}
@{}}
\toprule
\textbf{Paper} & \textbf{Semantic Target} & \textbf{Role of LLM} & \textbf{Downstream Task} & \textbf{Evaluation} \\
\midrule

\cite{luo2024deciphering}
& Trajectory semantic inference (occupation, activity sequence, etc)
& Chain-of-Thought reasoning
& No downstream task
& No quantitative metrics \\\midrule

\cite{ma2025learning}
& Prompting POI semantics for activity annotation
& Potential POI category recommender
& Trajectory data enrichment, POI classification
& JSD, accuracy, F1 score \\\midrule

\cite{li2025understanding}
& Identification of multimodal travel patterns
& Semantic embedding of multimodal travel features
& Semantic similarity measurement (clustering)
& Precision, recall, F1 score \\\midrule

\cite{chen2025enhancing}
& Semantic representation learning of locations
& Semantic encoding, instruction tuning
& Next-location prediction, mobility recovery, location alignment
& Hit@1, Hit@5, Hit@10 \\\midrule

\cite{ji2024evaluating}
& Measurement of trajectory distances
& Trajectory embedding extractor
& Destination prediction, distance measurement
& Correlation with classical distance metrics \\\midrule

\cite{yan2025valuing}
& Understanding human value of travel time
& Behavioral proxy reasoner for ranking alternatives
& Zero-shot prompting
& VOT elasticity, estimation error \\\midrule

\cite{li2024limp}
& Reasoning of latent mobility intentions
& Generation of intent distributions
& Mobility prediction
& Acc@1, Acc@5, Acc@10, MRR@5 \\\midrule

\cite{nie2025joint}
& Regional semantic understanding, geospatial semantic encoding
& Semantic region encoder
& Regional demand prediction
& MAE, RMSE \\
\bottomrule\bottomrule
\end{tabularx}
\end{table*}

\subsection{LLMs for mobility semantics and understanding} 
Human mobility semantics and understanding study how to extract \emph{meaning} from mobility data. Existing mobility data contains \emph{where} and \emph{when} people travel, many applications require additional semantic information such as \emph{what} people are doing (activity type), \emph{why} they move (intention), and \emph{how} locations function (home/work/leisure, etc.). Recent work leverages LLMs as semantic engines to infer these missing attributes, to learn semantically meaningful representations of trajectories and locations, and to assess whether LLMs can reproduce human preference trade-offs in mobility.

The first line of work uses LLMs to infer semantic attributes that are not explicitly available in raw mobility data. \citet{luo2024deciphering} formalize trajectory semantic inference along three outputs (i.e., occupation category, activity sequence, and narrative description) and propose TSI-LLM, combining spatiotemporal enhanced formatting (STFormat) with context-inclusive prompts to improve semantic interpretation from trajectory chains. \citet{ma2025learning} pursue a related goal from a data-quality perspective: because GPS traces and POI metadata are often incomplete or ambiguous, an LLM is used to map weak place descriptions into standardized activity semantics (with probabilities), which are then used to annotate stay points and support subsequent cross-domain fusion and trajectory completion.

The second line learns representations that capture semantic similarity beyond hand-crafted features. \citet{li2025understanding} convert multimodal trajectories into textual descriptions, derive BERT-based semantic embeddings, and apply DBSCAN clustering to identify interpretable multimodal travel patterns. \citet{chen2025enhancing} address a key bottleneck in LLM-based mobility modeling, i.e., location identifiers are discrete and semantically weak. They introduce semantic location tokenization via hierarchical vector quantization, representing each location as a short token sequence and instruction-tuning LLMs to better capture mobility semantics. The resulting representation supports tasks such as next-location prediction, mobility recovery, and location alignment. \citet{ji2024evaluating} evaluate trajectory embeddings derived from GPT-J and show that cosine distances in embedding space correlate strongly with classical trajectory distances (e.g., Hausdorff and DTW), while also emphasizing practical difficulties when tasks require accurate numerical reasoning or reliable identification of spatial neighbors. Mobility understanding also entails assessing whether LLMs reflect established behavioral trade-offs in transport. \citet{yan2025valuing} estimate value of travel time (VOT) from LLM responses to stated-preference ranking tasks under different contexts (e.g., income and trip purpose) using a rank-ordered logit analysis. Their results show meaningful behavioral similarities with human benchmarks at an aggregate level, but also model-dependent differences in sensitivity magnitudes and elasticity patterns, motivating careful validation before using LLMs as behavioral proxies.

Finally, semantic inference is often used as an intermediate layer that strengthens downstream mobility models. \citet{li2024limp} infer intention annotations (e.g., working, errands, leisure) from trajectories and represent them as an intent distribution that is added to a next-POI prediction model, improving performance by making behavioral states explicit. At a regional scale, \citet{nie2025joint} adopt an LLM as a geospatial encoder to extract semantic region representations from unstructured location data and build a semantics-informed graph to support inductive prediction across regions. Table \ref{table:llm_semantics_mobility} summarizes the related works.

\begin{table}[!htbp]
    \centering
    \caption{Comprehensive Comparison of Mobility Aspects and LLM Integration}
    \label{tab:mobility_comprehensive}
    \tiny
    \renewcommand{\arraystretch}{0.9}
    \makebox[\textwidth][c]{%
    \resizebox{1.2\textwidth}{!}{
    \begin{tabularx}{\textwidth}{@{} 
        >{\bfseries\raggedright\arraybackslash}p{2.2cm} 
        >{\raggedright\arraybackslash}X 
        >{\raggedright\arraybackslash}X 
        >{\raggedright\arraybackslash}X 
        >{\raggedright\arraybackslash}X 
        >{\raggedright\arraybackslash}X 
    @{}}
    
    \toprule
    \textbf{Aspect} & 
    \textbf{Itinerary Planning} & 
    \textbf{Trajectory Generation} & 
    \textbf{Mobility Simulation} & 
    \textbf{Mobility Prediction} & 
    \textbf{Semantics \& Understanding} \\
    \midrule

    Primary Objective & 
    Produce one itinerary for one traveler & 
    Produce realistic diaries/trajectories for a person or population & 
    Simulate system outcomes according to agents' interaction & 
    Predict a future variable (next location, mode, ETA, demand, etc.) & 
    Extract ``meaning'' from trajectories and semantic learning \\
    \midrule
    
    Output Type & 
    Time-ordered itinerary (often multi-day) & 
    Activity diary (what-when-where-how) or trajectory (locations over time) & 
    Realized trajectories with system states over time (congestion, travel times, accessibility) & 
    Predicted label/value (top-$k$ location, mode class, ETA, flow, etc.) & 
    Semantic labels (activity type, intent, etc.) or semantic embeddings/descriptions \\
    \midrule

    Primary Evaluation Target & 
    Constraint feasibility + factual grounding & 
    Realism, diversity, long-horizon coherence & 
    Network feasibility + supply-demand interactions + feedback dynamics & 
    {Predictive accuracy + robustness/transfer} & 
    Semantic plausibility, consistency, and agreement with annotations \\
    \midrule

    Core Difficulty & 
    Diverse constraints + evolving preferences + open-world facts & 
    Long-horizon realism + heterogeneity + missing intention in raw trajectories & 
    Contextual behavior modeling + replanning under disruptions + scalable interactions & 
    Domain shift (new cities/POIs), sparse context (rare events), class imbalance & 
    Semantic thinness/ambiguity, missing/noisy metadata, weak supervision \\
    \midrule

    Typical Traditional Baseline & 
    Rule-based itinerary planner; CP/MILP/heuristics & 
    Activity-based simulators; Markov/sequence models; deep generative models & 
    ABM/activity-based demand, traffic assignment and microsimulation (SUMO etc.) & 
    Spatiotemporal forecasting models; discrete choice/ML classifiers & 
    Rule-based labeling, probabilistic models, representation learning without LLM semantics \\
    \midrule

    LLM’s Main Role & 
    Intent/constraint extraction + plan drafting/repair + tool orchestration (with verifier/solver checks) & 
    Semantic enrichment + persona-conditioned synthesis (interpretable routines, purposes) & 
    Agent decision module in a closed loop (state/context $\to$ action/replanning; simulator execution) & 
    LLM-derived text embeddings/semantic features; RAG prompting for decisions & 
    Semantic inference/annotation (purpose, intent, etc.), semantic representation and descriptions \\
    \midrule

    Key Benefit of LLMs & 
    Makes planning usable from free-form requests; supports interactive revision for improving feasibility & 
    Adds meaning and controllability beyond coordinate-only generation & 
    Reduces hand-crafted behavior logic; enables contextual adaptation with network realism & 
    Improves transferability by injecting semantic meaning; improves robustness under sparse context & 
    Discovers interpretable semantics; improves downstream modeling and analysis \\
    \midrule

    Typical Evaluation & 
    Constraint satisfaction (pass rate), preference alignment, factual correctness & 
    Multiple realism metrics (temporal/spatial stats, transitions), diversity, coherence & 
    System metrics (congestion, accessibility, etc.), behavioral validity/stability & 
    Accuracy/F1/top-$k$, RMSE/MAE (ETA), cross-city generalization & 
    Agreement with labels/ontologies, consistency and interpretability; behavioral fidelity \\
    \bottomrule

    \end{tabularx}}}
    \label{comparationtable}
\end{table}

\section{Discussion}
\label{sec:dicsuccion}
\begin{figure}
    \centering
    \includegraphics[width=1.0\linewidth]{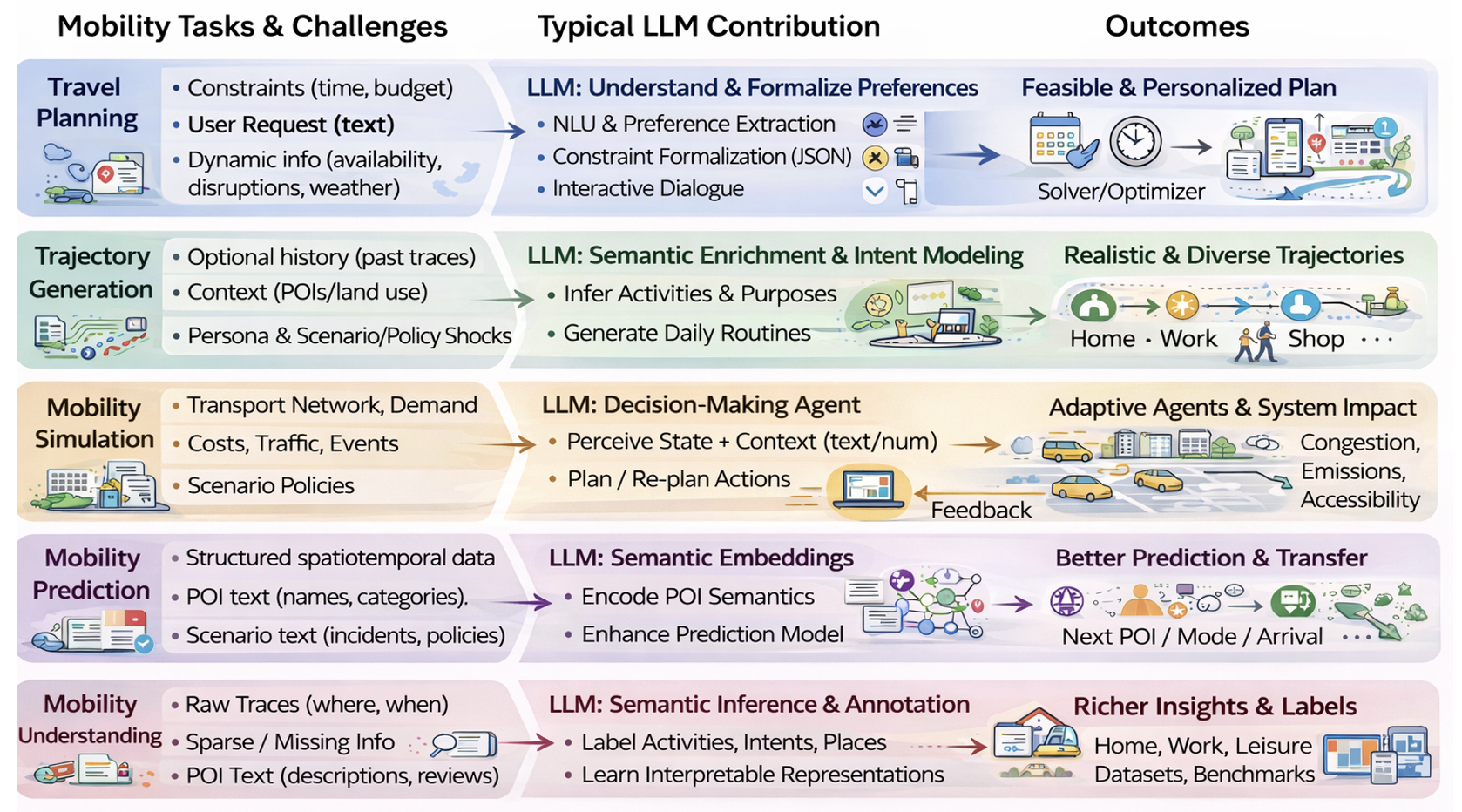}
    \caption{The typical opportunities and roles of LLMs for different human mobility tasks}
    \Description{This picture summarizes typical opportunities and roles of LLMs for different human mobility tasks.}
    \label{fig:discuss}
\end{figure}

The literature reviewed above shows that the popularity and effectiveness of LLMs across the core tasks of human mobility, including travel itinerary planning, trajectory generation, mobility simulation, mobility prediction, and understanding, stem from their unique ability to bridge the gap between raw numerical data and natural-language-based semantic information. Unlike traditional models that rely solely on numerical attributes (e.g., coordinates and time windows), LLM-based methods are better suited to processing subjective information that is difficult to quantify, such as intentions, preferences, and habits.  In the following, we (i) discuss task-specific challenges and how LLMs are used in each task, and (ii) synthesize unique roles of LLMs across tasks. Table~\ref{comparationtable} provides a comprehensive comparison across different aspects for those five tasks.  Fig. \ref{fig:discuss} provides a visual view of the typical opportunities and roles of LLMs for different human mobility tasks.

\subsection{Connecting task specific challenges with LLM opportunities}

\textit{Travel planning} is a challenging human mobility task as it requires coordinating many linked decisions under multiple constraints, while also capturing personal preferences that are often implicit. A feasible travel itinerary must respect feasibility constraints, e.g., opening hours, travel times, transfer times, reservation times, as well as overall limits such as budget and the number of nights. Meanwhile, travelers often articulate their preferences in informal and qualitative terms, such as “not too rushed”, “more local food”, or “I prefer visiting museums.” These preferences can even change in real time. Additionally, planning is also open-world, which means key information (e.g., availability, travel times, disruptions, weather) is distributed across different sources and may change over time. This causes: 1) purely rule-based systems require constant manual updates to accommodate new situations and are difficult to personalize at scale, and 2) purely predictive/recommendation models, which can rank POIs or predict next visits, are limited to translate free-form requests and unstructured texts (such as reviews) into explicit constraints. In contrast, LLMs are well suited due to their ability to perform semantic parsing and preference inference. They can parse underspecified natural-language requests, elicit missing requirements, and represent constraints and soft preferences (e.g., pace, interests, accessibility, vibe) in structured forms such as JSON.  Also, they can leverage semantically rich but unstructured information (e.g., POI descriptions and real user reviews) to infer implicit tastes and produce candidate plans aligned with latent preferences, as emphasized by preference-driven LLM-solver systems built on review-augmented benchmarks. LLMs can also be used for interactive specification and revision, meaning the model can ask targeted follow-up questions, propose a draft plan, and then edit the plan when the user changes preferences or when conflicts appear. To achieve reliability, LLMs  can be integrated into hybrid systems where the LLM handles natural language understanding and preference modeling, while verifiers and planners/solvers (e.g., SMT, CP, or MILP) validate feasibility and drive iterative repair when constraints conflict. 

\textit{Trajectory generation} is challenging because human mobility is sequential and context-dependent, and many travel decisions are driven by implicit purposes that raw data (e.g., GPS locations) do not reveal. Traditional approaches, such as rule-based activity simulators and data-driven generative models, can reproduce parts of observed mobility statistics. However, they often struggle to generate long-horizon traces that remain semantically consistent (e.g., maintaining plausible activity purposes and routines) and to generalize to new contexts where mobility patterns change (e.g., due to policy-driven mobility restrictions). LLM-based approaches are attractive alternatives due to their ability to perform semantic enrichment or intent inference over mobility data. By enriching locations with their semantic meaning (e.g., POI categories, reviews and other textual descriptions), an LLM may infer potential human intents and generate interpretable and plausible daily schedules. In addition, LLMs can be conditioned on persona and scenario descriptions (e.g., job type, preference, policy shocks) to produce controllable behavior hypotheses that are useful for simulation and stress-testing, although such outputs still require grounding and validation to ensure realism and avoid unjustified common-sense assumptions. 

\textit{Mobility simulation} typically models how many agents travel through a transport network and how their choices collectively create system-level outcomes such as congestion, travel times, and accessibility. Traditional ABM and activity-based simulation needs many explicit behavioral models (for activity scheduling, destination choice, mode choice, departure time choice, replanning rules). Building and calibrating these is time-consuming and usually depends on massive survey data and strong assumptions. As a result, LLMs are introduced as decision-making modules, which can 1) integrate heterogeneous inputs, such as numbers (e.g., travel times/congestion/costs),  texts (e.g., incident descriptions/policy notes), and turn them into a structured decision; 2) revise plans to be more coherent across linked choices (e.g., if a closure happens, an LLM may adjust departure time and reorder/drop activities from a relatively global view); and 3) generate behavior conditioned on persona and scenario descriptions without hand-coding a large set of special cases. 
In an interactive manner, the LLM contributes to flexible context-based decision-making and the simulator still enforces systematic feasibility and produces feedback to LLM for decision update.

\textit{Mobility predictions}, such as next location prediction, travel mode choice prediction, and arrival time estimation, rely on spatial-temporal information, but performance and transferability often depend on richer contexts than mere coordinates and time steps. This is because human movement is inherently driven by implicit motivations, intentions, and preferences that numerical data alone cannot fully capture. Traditional approaches can achieve strong accuracy within a fixed city and dataset, yet they commonly face generalization limitations because the learned embeddings are tied to city-specific features without considering semantic attributes of places and trips that could be generalizable across cities.
Instead, LLMs provide semantic representations from descriptions associated with mobility entities (such as POI names, categories, and reviews) that encode POI functionality and relevant activity. Subsequently, the predictor can leverage these transferable semantic representations instead of merely memorizing ID-level transitions. 

\textit{Mobility understanding} involves uncovering semantic information behind raw movements (such as human intentions, travel habits). This task is challenging since mobility data is typically semantically thin: trajectories encode where and when humans perform certain activities, but what analysts care about is more semantic information, such as activity type, trip purpose, intention, POI functionality, lifestyle context, which are latent and often unobserved. Meanwhile, the mapping from travel patterns to semantic meaning is many-to-many: the same movement can correspond to different purposes (a stop near an office could be for work, meeting, delivery, etc), and the same purpose can appear in many patterns, e.g., following individual habits and external constraints. Data noise, sparsity, incompletion (such as missing POI metadata), and cross-city differences make this task much harder. LLMs are used for mobility understanding because they can act as a text-conditioned semantic inference module. In this regard, an LLM is primarily doing inference and representation extraction, by which it tries to recover missing data (e.g., attributes, labels) or learn embeddings that convey semantic meaning. As an intermediate layer, the output of LLM can subsequently be used for prediction, generation, or simulation.

\subsection{Task specific roles of LLMs in human mobility}
Across tasks, LLMs are not replacing mobility models in a single uniform way. However, their role differs by task. For travel planning, the primary benefit of LLMs is their ability to act as a symbolic translator that bridges the gap between unstructured human desires and hard mathematical constraints. For diary/trajectory generation, the goal is not a single feasible plan, but realistic and diverse synthetic mobility behavior over long horizons. Here LLMs are attractive because they can incorporate semantic and persona-level conditioning that is hard to encode in conventional generators. For mobility simulation, the LLM is mainly an agent decision module inside an environment. It takes the simulator’s state (and possibly unstructured scenario inputs) and proposes actions or (re)planning decisions. The simulator executes them and returns feedback, enabling context-dependent adaptation and richer agent logic, while the simulator remains the source of truth for feasibility and supply-demand interactions. For mobility prediction, LLMs mainly work on converting textual information associated with mobility entities, such as POI names, categories, and descriptions, into LLM-derived text embeddings (e.g., POI semantic embeddings) that encode POI semantics in a transferable way. These embeddings can be fused with spatiotemporal features, so the predictor relies less on memorizing ID transitions and more on functionally meaningful similarity across places. For mobility understanding, LLMs are used to infer and represent the “what/why” behind mobility traces. This includes annotating trajectories with activity and intention labels, inferring functional roles of locations, learning their semantic representations that support clustering and interpretability, and building benchmarks that test whether models can answer semantic questions about trajectories.

\section{Open Challenges and Future Directions}
\label{sec:challenge}
After connecting task-specific challenges to the ways LLMs are used, this section summarizes LLMs limitations that repeatedly exist across tasks, and outlines a future research agenda.

\paragraph{Feasibility with real travel times and constraints} A recurring limitation is that LLMs often produce itineraries or mobility traces that appear reasonable on the surface, but once real-world constraints are checked, the output becomes infeasible. In travel itinerary planning, common failures include visiting places outside opening hours, exceeding the stated budget, violating time windows, or assuming unrealistically short travel times between locations. Recent benchmarks therefore evaluate these violations directly by measuring how often a generated itinerary satisfies all hard constraints~\citep{chen2024travelagent,chen2024can}. Similar feasibility problems appear in trajectory generation and mobility simulation. An LLM agent may describe a seemingly plausible sequence of activities (e.g., from home to work to shopping), yet the implied movements can still be impossible: the trajectory may require “teleporting” across long distances, pack too many activities into too short time, or select actions that cannot be executed on a transport network once travel times and capacity effects are considered. A common solution is a hybrid design in which the LLM proposes plans while external modules compute travel times, verify constraints, and repair violations. Even though effective, this shifts the core research question from “can the LLM plan” to “can the system guarantee feasibility and, at the same time preserving intent.” This leads to two open issues. First, solver and verifier-in-the-loop designs increase system complexity and latency, especially in interactive settings with frequent user revisions. Second, repairs should be preference-preserving which means fixing infeasibility should not silently replace the user’s priorities with a qualitatively different plan.

\paragraph{Dynamics and interaction effects} Most current evaluations assume a snapshot world, where travel times, prices, and availability are fixed during planning. In practice, mobility decisions are inherently dynamic where hotels sell out, ticket prices change, disruptions occur, preferences change over time and multiple users compete for the same limited capacity. Such dynamics become more challenging in multi-user settings because competition for scarce resources (rooms, tickets, peak-hour capacity) and network feedback (congestion and crowding) make outcomes depend on other travelers’ actions rather than on a fixed background state.
Even in mobility simulation, many LLM-agent setups still model travelers in isolation, which makes it difficult to reproduce interaction-driven macro outcomes (e.g., crowding, congestion, and demand shifting) that ABMs are designed to capture. Recent work therefore argues that realism at the aggregate level requires moving beyond single-agent behavior toward explicit multi-agent interaction~\citep{wang2024large}. Addressing the snapshot-world limitation requires evaluation protocols and models that incorporate time-varying information and capacity constraints, e.g., dynamic evaluation protocols that incorporate time-varying updates (e.g., availability, schedule changes, disruptions) and explicit capacity/booking rules. Agents can be scored by their replanning quality, e.g., whether they can adapt with minimal unnecessary changes while preserving user priorities. Addressing the interaction limitation requires explicit mechanisms for multi-user effects, such as congestion/crowding feedback and coupled decisions under shared constraints (e.g., households, group travel, or social coordination).

\paragraph{Semantic and spatiotemporal representation gap}
LLMs are strong at extracting and reasoning over semantic context (e.g., POI descriptions, activity purposes, and narrative constraints), but they are not naturally reliable at representing quantitative spatiotemporal structure. As a result, in mobility prediction, LLMs are rarely used as standalone predictors. Most work instead rely on retrieval, intent inference, hierarchical structures, and downstream statistical or learning-based models to capture the underlying spatiotemporal relationships~\citep{wang2023would,liang2024exploring}. Many approaches also reformulate trajectories or numerical attributes in texts, which can discard geometric information and increase sensitivity to prompt design, reducing robustness across datasets and scenarios~\citep{xu2025evaluating,li2025zero}. A similar limitation appears in mobility understanding: semantic embeddings can improve interpretability but still fail to encode key spatiotemporal relationships such as distance, proximity, and movement continuity~\citep{li2025understanding}. Future work should therefore combine semantic representations with explicit spatial structure (coordinates, networks, and graphs) and evaluate both semantic plausibility and spatiotemporal consistency. When LLMs are used to infer behavioral quantities (e.g., preferences or value of travel time), outputs should be calibrated and validated against empirical benchmarks rather than treated as substitutes by default.
\paragraph{Evaluation comparability, validation, and reproducibility} Across tasks, a central challenge is to ensure that reported improvements are comparable and robust, rather than driven by a particular choice of metrics or experimental settings. Current evaluation metrics are fragmented, which makes it hard to compare methods and easy to overlook unintended trade-offs. In trajectory generation, studies often report different subsets of realism metrics as shown in Table \ref{table:llm_trajectory}. As a result, a method may appear better on one dimension while quietly getting worse on another, such as matching routine regularity more closely but producing less realistic spatial dispersion or step-distance patterns.
Beyond comparability, validation and robustness raise additional difficulties. In mobility simulation, behavior that looks plausible at the individual level does not necessarily reproduce correct aggregate outcomes once network interactions and feedback are taken into account. In addition, reported results can vary with prompts, model versions, and sampling settings, raising reproducibility concerns. Recent work highlights that behavioral alignment is multi-dimensional and calls for validation datasets and tests that evaluate both distributional realism and decision consistency under controlled perturbations~\citep{liu2025toward}. For mobility understanding, QA-style benchmarks further suggest that factual retrieval can be relatively strong while semantic reasoning and long-horizon explanation remain difficult~\citep{asano2025mobqa}. Overall, stronger community standards are needed, including unified metric suites, controlled ablations, robustness checks, and transparent reporting of sensitivity to prompting and decoding.

\paragraph{Scalability and computational cost}
LLM-based approaches to human mobility can be costly at city scale, especially in simulation where each decision step may require an LLM inference call. This limits practical deployment and makes large-scale calibration and uncertainty quantification difficult. Reducing the number of LLM calls without sacrificing behavioral fidelity remains an open research and engineering challenge. A promising near-term direction is hybrid modeling: use LLM agents primarily for high-level decisions that are hard to formalize, and delegate structured components, such as routing, network loading, and mode/route choice (when appropriate), to established transportation models.

\paragraph{Privacy risks} 
Synthetic mobility data is often presented as a privacy-preserving alternative to sharing raw trajectories, but high data usefulness typically requires high fidelity, which can still enable privacy attacks. This concern is amplified in mobility because trajectories are highly unique, making rare or distinctive movement patterns especially easy to associate with individuals. Privacy therefore should be treated as an evaluated property, rather than an assumed benefit. Future work should design privacy and usefulness related measures. 

In summary, we expect future progress to come less from increasingly complex prompting and more from building systems whose outputs can be verified and assessed in a transparent way. One direction is to couple LLMs with explicit checking components (e.g., routing tools and constraint checkers) so that hard requirements such as opening hours, time windows, budgets, and travel time feasibility are verified automatically. When violations occur, revisions should preserve the user’s key priorities or preferences rather than produce a qualitatively different plan. A second direction is to develop benchmarks that reflect real planning as an iterative process, where new information arrives over time (e.g., disruptions, delays, changing prices, sold-out options) and agents are evaluated by how well they update plans with minimal unnecessary changes. Such benchmarks could later be extended to include competition among multiple users for limited capacity. A third direction is to adopt shared evaluation protocols with a common set of metrics covering feasibility, spatiotemporal realism, robustness to small input or prompt changes, and privacy risk versus data usefulness, so that the results across studies are directly comparable. Finally, enabling city-scale deployment will require computational efficiency oriented designs that reduce the number of LLM calls (e.g., caching previously computed results, limiting calls to decision points that truly require language reasoning, or distilling a large model into a smaller one) while maintaining behavioral realism.


\section{Conclusion}
This survey reviews recent progress in applying large language models to human mobility research. Although a growing body of work exists in this area, we focus on five main directions: LLMs for travel itinerary planning, LLMs for trajectory generation, LLMs for mobility simulation, LLM for mobility prediction, and LLM for mobility semantics and understanding. According to the comprehensive literature review, we summarize and discuss LLM opportunities and roles in human mobility tasks. The challenges faced by current methods are analyzed, thereafter highlighting several promising future research directions. We hope this survey can inspire further studies to address these challenges and advance LLM-based approaches in human mobility.

\bibliographystyle{ACM-Reference-Format}
\bibliography{sample-base}
\end{document}